\documentclass{aa}
\usepackage{graphics,natbib}

\begin{document}

\title{Anomalous X-ray line ratios in the cTTS TW\,Hya}

\author{J.-U. Ness$^1$\and J.H.M.M. Schmitt$^2$}
\institute{
Department of Physics, Rudolf Peierls Centre for Theoretical Physics,
University of Oxford, 1 Keble Road, Oxford OX1\,3NP, UK
\and
Hamburger Sternwarte, Universit\" at Hamburg, Gojenbergsweg 112,
D-21029 Hamburg, Germany}

\authorrunning{Ness and Schmitt}
\titlerunning{Anomalous X-ray line ratios in the cTTS TW\,Hya}

\offprints{J.-U. Ness}
\mail{ness@thphys.ox.ac.uk}
\date{Received \today; accepted soon...}

\abstract{The cTTS TW\,Hya has been observed with high-resolution X-ray
spectrometers. Previously found high densities inferred from He-like f/i
triplets strongly suggested the detected X-ray emission to be dominated by an accretion
shock. Because of their radiation field dependence He-like f/i ratios do not provide
unambiguous density diagnostics. Here we present additional evidence for high densities
from ratios of Fe\,{\sc xvii} lines. Key Fe\,{\sc xvii} line ratios in TW\,Hya
deviate from theoretical expectations at low densities as well as from the
same measurements in a large sample of stellar coronae. However, a quantitative
assessment of
densities is difficult because of atomic physics uncertainties. In addition, estimates
of low optical depth in line ratios sensitive to resonance scattering effects also
support a high-density emission scenario in the X-ray emitting regions of cTTS.
}

\maketitle

\section{Introduction}

T Tauri stars are young pre-main sequence (PMS) late-type stars. ``Classical" T Tauri
stars (cTTS) are thought to still be surrounded by accretion disks as evidenced by
IR and UV excess, while no signs for the presence of a disk are found in the so-called
``weak line'' T Tauri stars \citep[for details we refer to][]{feigmon99}. 
X-ray emission from PMS stars is expected to be high because of their fast
rotation if the emission is interpreted as scaled-up solar-type activity. However, for
cTTS an additional source of X-ray emission through accretion is available; this
additional X-ray production mechanism is expected to lead to significant differences
in X-ray emission levels and variability, but in particular to differences in the
spectral properties of the X-ray emission.\\
High-resolution X-ray observations with the
transmission and reflection gratings aboard {\it Chandra} and {\it XMM-Newton} have
now been obtained for about two dozens of stars, but only for very few cTTS. The best
data are usually obtained for the O\,{\sc vii} triplet located at 21.6\,\AA\ (``r-line''),
21.8\,\AA\ (``i-line''), and 22.1\,\AA\ (``f-line''). The f/i-line ratio is
density-sensitive \citep{gj69}, but in all cases no ratios below unity are encountered
for coronal sources \citep{denspaper}. In contrast, the available high-resolution
spectra of cTTS show unusually low He-like f/i ratios in TW\,Hya
\citep{kastner02,stelz04} and BP Tau \citep{bptau}. The first obvious
conclusion was that the plasma in cTTS is at extremely high densities
suggesting its origin in an accretion shock rather than a ``normal'' magnetically active
corona \citep{kastner02}. However, f/i ratios could only be measured for O\,{\sc vii}
and Ne\,{\sc ix}, and the f/i ratios of those ions also depend on UV radiation
fields if they are strong enough and located close to the origin of the X-ray emission.
Since the presumed accretion shock region is also expected to produce
intense UV emission, the observed low f/i-ratios would then not contradict the
accretion hypothesis, but need not necessarily imply high densities.\\
The {\it Chandra} HETGS spectrum of TW\,Hya has already been analyzed by
\cite{kastner02} with a variable abundance differential emission measure
analysis and the identification and discussion of the anomalously low
f/i-ratios in O\,{\sc vii} and Ne\,{\sc ix}. We address an alternative approach to
density determination using Fe\,{\sc xvii} lines at 17.05\,\AA\ and 17.10\,\AA.
\cite{mauche01} were the first to use this ratio as density tracer
in their study of the {\it Chandra} HETGS spectrum of the intermediate
polar EX\,Hya, demonstrating that this ratio is considerably less sensitive to
photoexcitation than He-like ions. Further, we investigate the effects of resonant
line scattering which also depends sensitively on plasma density.\\
The atomic physics of Fe\,{\sc xvii}, especially the $3d\rightarrow2p$
(15\,\AA\ range) and $3s\rightarrow 2p$ (17\,\AA\ range) transitions, is quite
complicated and extensive efforts have been spent with the conclusion that
a number
of indirect processes have to be considered apart from the standard collisional
excitation (CE) theory. The inclusion of resonance excitation and inner-shell
excitation from Fe\,{\sc xvi} as well as radiative and dielectronic recombination
from Fe\,{\sc xviii} improved the situation enormously, but residual discrepancies
remain \citep[for more
details see][]{gu03a}. In view of these difficulties we focus our analyses on the
comparison of TW\,Hya with a sample of stellar coronae, but refrain from any
quantitative determination of densities. Comparison with theoretical calculations
is based on atomic data using APEC \citep{smith01}\footnote{Version 1.3.1; available at
http://cxc.harvard.edu/atomdb} and the most recent calculations by \cite{gu03a}.
\begin{figure*}[!ht]
\resizebox{\hsize}{!}{\includegraphics{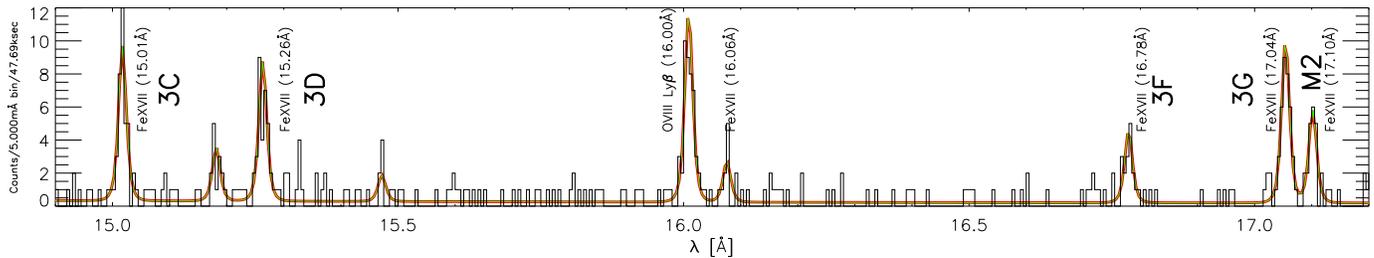}}
\caption{\label{fe17}HETGS spectrum and a parameterized best-fit model of TW\,Hya
(summed plus- and minus orders); line letter designations 
refer to the line parameters given in Tab. \ref{tab1}.
}
\end{figure*}

\vspace{-.5cm}
\section{Data Reduction and Analysis}

{\it Chandra} observations of TW\,Hya (K7\,Ve, $d=57$\,pc) were carried out
with the High Energy Transmission Grating Spectrometer (HETGS; ObsID 5, 48\,ksec,
June 2000); details are given by \cite{kastner02}.
We extracted count spectra using the Chandra Interactive Analysis of Observations (CIAO)
software, but used our own tool CORA \citep{newi02} to determine line counts by
fitting line templates and converted them to line fluxes using effective areas extracted
with CIAO (using the tool {\tt fullgarf}). After examining the spectra of the plus and
minus sides we use the sum (representing the effective area-weighted average) for our
analysis, which can be done if no anomalies occur on either side (which is not the
case to our knowledge). Any correction of line fluxes for absorption is based
on a value of $N_{\rm H}=4\times 10^{20}$\,cm$^{-2}$, derived from a
broad-band spectrum of TW\,Hya by Robrade et al. (2005).
For comparison we extrated the HETGS spectra of various stellar
coronae in exactly the same way to assess
spectral differences between TW\,Hya and purely coronal sources.
\begin{table}[!ht]
\begin{flushleft}
\renewcommand{\arraystretch}{1.1}
\caption{\label{tab1}Line flux measurements for TW\,Hya.}
\vspace{-.6cm}
\begin{tabular}{p{.1cm}p{.7cm}p{1.3cm}p{.4cm}cp{.6cm}c}
\hline
&$\lambda$ [\AA] &\ \ \ counts & A$_{\rm eff}^{\rm [cm^2]}$ & ion & transm.$^{[a]}$&flux$^{[b]}$\\
3C& 15.01 & \mbox{$35.5\,\pm\,6.2$} & 16.0 & Fe\,{\sc xvii} & 0.859 & $7.15\,\pm\,1.26$\\
3D& 15.26 & \mbox{$32.3\,\pm\,6.0$} & 23.2 & Fe\,{\sc xvii} & 0.853 & $4.45\,\pm\,0.83$\\
\multicolumn{2}{r}{$^{[c]}$16.00} & \multicolumn{2}{l}{blend in O\,{\sc viii}} & Fe\,{\sc xviii} & 0.835 & $[1.15\,\pm\,0.41]^{[c]}$\\
\multicolumn{2}{r}{$^{[c]}$16.07} & \mbox{$9.39\,\pm\,3.4$} & 19.0 & Fe\,{\sc xviii} & 0.834 & $1.53\,\pm\,0.55$\\
3F& 16.78 & \mbox{$16.6\,\pm\,4.4$} & 16.1 & Fe\,{\sc xvii} & 0.816 & $3.13\,\pm\,0.83$\\
3G& 17.05 & \mbox{$37.6\,\pm\,6.4$} & 14.8 & Fe\,{\sc xvii} & 0.809 & $7.64\,\pm\,1.30$\\
M2& 17.10 & \mbox{$21.2\,\pm\,4.8$} & 14.6 & Fe\,{\sc xvii} & 0.807 & $4.36\,\pm\,1.00$\\
&10.23 & \mbox{$45.0\,\pm\,7.3$} & 83.7 & Ne\,{\sc x} & -- & $2.18\,\pm\,0.35$\\
&12.13 & \mbox{$200\,\pm\,14.5$} & 43.1 & Ne\,{\sc x} & -- & $15.8\,\pm\,1.15$\\
&16.00 & \mbox{$47.3\,\pm\,7.1$} & 19.2 & O\,{\sc viii} & -- & $6.39\,\pm\,0.97$\\
&18.97 & \mbox{$121\,\pm\,11.2$} & 8.87 & O\,{\sc viii} & -- & $29.8\,\pm\,2.78$\\
\hline
\end{tabular}
$^{[a]}$Transmission efficiencies with $N_{\rm H}=4\times 10^{20}$\,cm$^{-2}$\\
$^{[b]}10^{-14}$\,erg\,cm$^{-2}$\,s$^{-1}$; corrected for $N_{\rm H}$\\
$^{[c]}\sim75$\% of 16.07-\AA\ line blends with O\,{\sc viii} at 16.00\,\AA\\
\renewcommand{\arraystretch}{1}
\end{flushleft}
\vspace{-.6cm}
\end{table}

\section{Results}

\begin{figure}[ht]
\resizebox{\hsize}{!}{\includegraphics{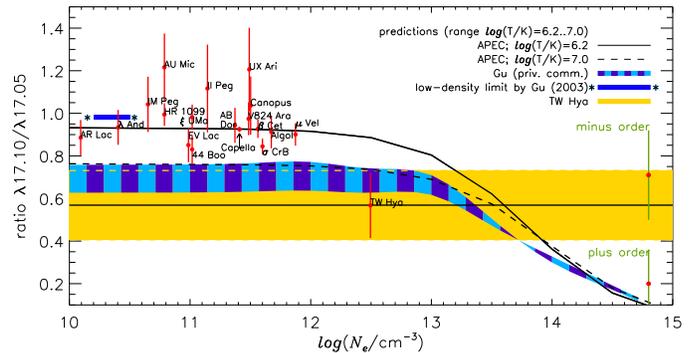}}
\caption{\label{fe1717}Density dependence of the Fe\,{\sc xvii} line ratio
$\lambda\lambda 17.10/17.05$ predicted by APEC and M.F. Gu compared with
18 coronal measurements and TW\,Hya; plasma densities are derived from
Ne\,{\sc ix} f/i ratios.}
\end{figure}

For our analyses we measured fluxes for two sets of lines, five lines of Fe\,{\sc xvii}
and the Ly$\alpha$ and Ly$\beta$ lines of
H-like oxygen and neon. Fig.~\ref{fe17} shows the HETGS spectrum between 15--17.2\,\AA\
illustrating the reliability of our line detections and flux measurements.
The best fit above a source continuum of 20\,cts/\AA\ is also shown
(FWHM 0.016\,\AA\ for all lines), and the derived counts and fluxes are listed in
Table~\ref{tab1} together with the effective area values used for conversion of
line counts to fluxes and the transmission efficiencies for $N_{\rm H}=4\times
10^{20}$\,cm$^{-2}$ calculated from \cite{bamabs}; we assume standard cosmic
abundances from \cite{agrev89}. The Fe\,{\sc xvii} line fluxes are corrected
for absorption, but we did not correct
the Ly$\alpha$ and Ly$\beta$ lines which will be investigated in detail below.
In Table~\ref{tab1} we also list an Fe\,{\sc xviii} line at 16.07\,\AA\
(2p$^4$3s$\,^2$P$_{5/2}$ to ground state) which we use to correct the O\,{\sc viii}
Ly$\beta$ line at 16.00\,\AA\ to account for contamination by Fe\,{\sc xviii} at
16.004\,\AA\ (2p$^4$3s$\,^2$P$_{3/2}$ to ground). The ratio of these two
Fe\,{\sc xviii} lines varies slowly with temperature from 0.73 to 1.06 for
$\log T=6.3-7.3$ (as predicted by APEC). Given the low temperature of
TW\,Hya, the measured flux in O\,{\sc viii} Ly$\beta$ was reduced by
75\% of the flux in the 16.07-\AA\ line. We corrected 
the stellar O\,{\sc viii} Ly$\beta$ line fluxes by interpolating the
slow temperature dependence using temperature estimates from the ratio of lines of
O\,{\sc viii} at 18.97\,\AA\ and O\,{\sc vii} at 21.6\,\AA. The line fluxes in
Table~\ref{tab1} are used to calculate the line flux ratios given in
Table~\ref{ratios}. For the Fe\,{\sc xvii} line ratios the $N_{\rm H}$-corrected
fluxes were used while the Ly$\beta$/Ly$\alpha$ ratios are (not yet) corrected;
the O\,{\sc viii} ratio is corrected for contamination with Fe\,{\sc xvii}, but
blending in the Ne\,{\sc x} Ly$\alpha$ line is ignored.\\
\begin{table}[!ht]
\begin{flushleft}
\renewcommand{\arraystretch}{1.1}
\caption{\label{ratios}Line flux ratios in TW\,Hya (notation as in Fig.~\ref{fe17}).}
\vspace{-.2cm}
\begin{tabular}{p{.9cm}p{1.4cm}r|p{.9cm}p{1.4cm}r}
\hline
$\lambda\lambda$ & flux ratio & av$^{[a]}$& $\lambda\lambda$ & flux ratio& av$^{[a]}$\\
M2/3G & \mbox{$0.57\,\pm\,0.16$} & 0.9 & 3D/3F & \mbox{$1.42\,\pm\,0.46$}&0.6\\
3G/3F & \mbox{$2.43\,\pm\,0.76$} & 1.3 & 3G/3C & \mbox{$1.06\,\pm\,0.26$}&0.8\\
M2/3F & \mbox{$1.39\,\pm\,0.48$} & 1.2 & 3G/3D & \mbox{$1.71\,\pm\,0.43$}&2.1\\
3D/3C & \mbox{$0.62\,\pm\,0.16$} & 0.4 & 3F/3C & \mbox{$0.43\,\pm\,0.19$}& 0.6\\
O($\beta/\alpha$) & \mbox{$0.18\,\pm\,0.03$} & 0.14 &Ne($\beta/\alpha$)& \mbox{$0.13\,\pm\,0.02$}&0.14\\
\hline
\end{tabular}
$^{[a]}$error-weighted average of stellar measurements
\renewcommand{\arraystretch}{1}
\end{flushleft}
\vspace{-.5cm}
\end{table}
We compare the line ratios of TW\,Hya with a large sample of analogous measurements of
stellar coronae in a variety of classes. We first focus on the ratio
$\lambda\lambda17.10/17.05$ as a function of density in Fig.~\ref{fe1717}; TW\,Hya is
indicated by the light shaded area denoting the $1\sigma$ uncertainty range from
measurement errors. Stellar coronal densities are calculated from Ne\,{\sc ix}
f/i ratios given for 18 stars by \cite{denspaper},
cleared of all the Fe\,{\sc xix} blending. A clear discrepancy between all stellar
measurements and TW\,Hya can be recognized. Theoretical predictions from APEC and
Gu (priv. comm.; not yet including indirect processes), all in the temperature range
$\log T=6.2-7.0$, are shown for two
temperatures bracketing those temperatures where the bulk of the Fe\,{\sc xvii} line
formation is expected to occur. While with APEC a quantitative determination of a
high density $\sim 4\times 10^{13}$\,cm$^{-3}$ is possible, the measurement of
TW\,Hya does not deviate from the low-density limit as predicted by Gu.
The stellar measurements appear more consistent with the APEC-prediction, however,
since all coronal sources are thought to be far hotter than $\log (T/{\rm K})=6.0$,
also APEC underpredicts the measured 17.10/17.05-ratios. In contrast, the calculations
by \cite{gu03a}, including indirect processes (but providing only the low-density limit
at different temperatures), agree very well with the stellar measurements and an
expansion of these calculations including all indirect processes as a function of density 
is likely to provide better estimates
for TW\,Hya. Obviously, the observed Fe\,{\sc xvii} 17.10/17.05-ratio in
TW\,Hya is smaller than that typically measured in coronal sources and it is
larger than in EX\,Hya \citep[cf. ][]{mauche01},
suggesting a plasma density in TW\,Hya larger than typically encountered in coronal
sources, but smaller than $2\times 10^{14}$ cm$^{-3}$ as inferred for EX\,Hya.
We also compare more ratios in Table~\ref{ratios} with our
stellar sample. In particular we found TW\,Hya to be different in the ratios
$\lambda\lambda17.05/16.78$ (3G/3F) and $\lambda\lambda15.26/15.01$
(3D/3C), while in the other ratios no differences between
TW\,Hya and the stellar sample are found. This suggests that the 17.05-\AA\ and the
15.26-\AA\ lines are anomalously enhanced in TW\,Hya while those at
15.01\,\AA, 16.78\,\AA, and 17.10\,\AA\ show no peculiarities.\\
%
A comparative analysis of the line ratios $\lambda\lambda15.26/15.01$ and (3D/3C)
$\lambda\lambda15.26/16.78$ (3D/3F) has been carried out by \cite{ness_opt} to
investigate opacity effects. For both ratios TW\,Hya shows significantly larger
ratios possibly indicating resonant line scattering effects. However, laboratory
experiments suggest blending of the 15.26-\AA\ line with an Fe\,{\sc xvi} satellite
line \citep[e.g.,][]{brown01}, possibly explaining a trend of increasing
$\lambda\lambda15.26/15.01$-ratios with decreasing temperature \citep{ness_opt}.
Since TW\,Hya does have
relatively low X-ray temperature \citep{kastner02,stelz04}, the same reason can account
for the anomalously high flux in the 15.26-\AA\ line. In view of this ambiguity
we also studied the H-like Ly$\beta$/Ly$\alpha$ line ratios of O\,{\sc viii}
and Ne\,{\sc x}. In Fig.~\ref{lyab} we show a sample of stellar ratios of O\,{\sc viii}
with the blending correction applied as described above, compared with the measurement
of TW\,Hya and atomic data predictions;
the same line ratio has been used by \cite{testa_opt} finding II\,Peg and IM\,Peg to
be anomalous. The TW\,Hya measurement is marked
with the hashed box bracketing the formal $1\sigma$ uncertainties in the line ratio
and in temperature. Theoretical predictions
from APEC and {\it Chianti} agree quite well with each other
and uncertainties in $N_{\rm H}$ do not lead to different conclusions.
However, we note in this context that the above blending correction is
not straightforward, since it is predicted rather differently
by {\it Chianti} and APEC (which we used for the correction).
In view of the stellar measurements and the fact that the error bars do not include
uncertainties from the blending correction,
this deviation from theory appears rather marginal. We carried out the same procedure
for the Ne\,{\sc x} lines and found no deviation from the stellar measurements
(Table~\ref{ratios}).

\vspace{-.4cm}
\section{Discussion and Conclusions}

\begin{figure}[t]
\resizebox{\hsize}{!}{\includegraphics{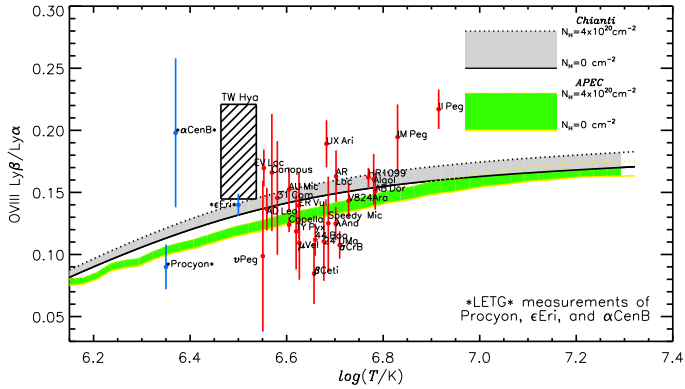}}
\caption{\label{lyab}Theoretical predictions of Ly$\beta$/Ly$\alpha$ ratio as
a function of temperature for O\,{\sc viii} and measurements of TW\,Hya
compared with stellar measurements (temperatures from
O\,{\sc viii}/O\,{\sc vii} line ratios); shaded areas are predictions for
values of $N_{\rm H}<4\times 10^{20}$\,cm$^{-2}$.}
\end{figure}

We identified and measured a new line flux ratio sensitive to density, which
strongly supports earlier conclusions of high plasma densities in TW\,Hya.
The ratio of $\lambda\lambda 17.10/17.05$ is a sensitive tracer of
high densities with little contamination from UV radiation or temperature.
Besides the extremely low f/i ratios in O\,{\sc vii} and Ne\,{\sc ix} we also
found this ratio to be anomalously low compared to all stellar coronae.
Unfortunately, quantitative constraints on density are still ambiguous because
theoretical calculations do not yet cover the full range of interactions between the
ground state and excited states. At any rate, the density of TW\,Hya appears to be
higher than that of typical stellar coronae, but lower than that of the intermediate
polar EX\,Hya. Unambiguous X-ray density measurements as
obtained here are also important for other cTTS.\\
The 15.01-\AA\ and 15.26-\AA\ Fe\,{\sc xvii} lines provide a sensitive test for the
effects of resonant line scattering. A comparison of this ratio for TW\,Hya with a sample
of cool stars \citep{ness_opt} shows a large value for TW\,Hya albeit with substantial
error; similarly large values are found for EV\,Lac and Prox\,Cen. However, as
discussed by \cite{ness_opt} the relative strength of the 15.26-\AA\ line can also be
explained by blending with low-temperature lines, leading us to conclude that the line
ratio of the 15.01-\AA\ and 15.26-\AA\ lines provides no unambiguous evidence for
resonance scattering in TW\,Hya. Next, we examined the ratios of H-like
Ly$\beta$/Ly$\alpha$ lines, which are also sensitive to resonant scattering (cf.
Fig.~\ref{lyab}). While the Ly$\beta$/Ly$\alpha$ line ratio for TW\,Hya is clearly
above the theoretical expectation, it does not differ significantly from those
encountered in stellar coronae. Also, there is a nagging uncertainty about the blending
correction with Fe\,{\sc xviii}, so that any resonant scattering effects appear
marginal. Since the situation is similar for Ne\,{\sc x} we conclude that there is
no clear evidence for any X-ray optical depth effects in TW\,Hya and that the optical
depth $\tau$ should be below unity.\\
%
\begin{figure}[ht]
\resizebox{\hsize}{!}{\includegraphics{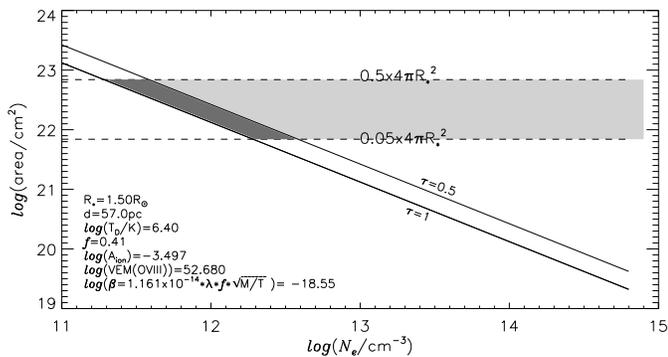}}
\caption{\label{ff}Area vs. density $N_e$ from the requirement
optical depth $\tau < 1$; area values for surface filling factors are indicated.}
\end{figure}
%
The apparent absence of resonant scattering and the value of the observed emission
measure in a line can be combined as follows: We use the formula
for the optical depth $\tau$ at line center derived by \cite{bhatsab01}
$\tau=1.61\times 10^{-14}f\lambda\sqrt{\frac{M}{T_D}} N_{\rm ion}L=\beta N_{\rm ion}L$
where $f$ denotes the oscillator strength and $\lambda$ wavelength, $M$ 
atomic mass, $T_D$ temperature, $N_{\rm ion}$ ion density, and $L$ path
length. Next, the product $N_{\rm ion} L $ can be expressed in terms of
the volume emission measure \mbox{VEM$ = N_e^2 L A$}, where $N_e$ denotes the electron
density and $A$ the area of the assumed cylindrical emission region. $N_e$ and
$N_{\rm ion}$ are related through $N_{\rm ion} = 0.85 N_e A_{\rm ion}$, where
$A_{\rm ion}$ denotes the abundance of the considered ion relative to hydrogen, 
assumed to be $3.18\times 10^{-4}$. We thus obtain \mbox{VEM$ = \tau \frac{N_e A}{0.85
A_{\rm ion} \beta}$}. In Fig. \ref{ff} we plot (assuming $\tau = 1$ and $0.5$)
$A$ as a function of $N_e$; note that this curve moves up for $\tau$-values below unity.
Since the area shown is bounded above by (half) the stellar surface, we also plot a
shaded area corresponding to surface filling factors of accretion hot spots between 0.5
and 0.05. If these filling factors are indeed of the order of a few percent as usually
assumed for cTTS, it is clear that densities in excess of $10^{12}$\,cm$^{-2}$ are
required to account for both the observed emission measure in TW\,Hya as well as the
absence of any clear optical depth effects.
\begin{acknowledgements}
We thank Prof. C. Jordan for sharing her profound experience with Fe\,{\sc xvii}
ensuring that our discussion is consistent with all important background information
of atomic physics and Dr. Brickhouse and Dr. Huenemoerder for discussion
of instrumental issues.
J.-U.N. acknowledges support from PPARC under grant number PPA/G/S/2003/00091.
\end{acknowledgements}
\vspace{-0.5cm}
\bibliographystyle{aa}
\bibliography{astron,jn,jhmm}

\end{document}